\documentclass[11pt]{article}

\usepackage{amsmath}
\usepackage{amssymb}
\usepackage{latexsym}
\usepackage{enumitem}
\usepackage{mathrsfs}
\usepackage{comment}
\usepackage{color}

\newtheorem{assumption}{Assumption}

\def\qed{ \ \vrule width.2cm height.2cm depth0cm\smallskip}
\newenvironment{proof}{\noindent {\bf Proof.\/}}{$\qed$\vskip 0.1in}

\newcommand{\ol}{\overline}
\newcommand{\ul}{\underline}

\newcommand{\ba}{\begin{array}}
\newcommand{\ea}{\end{array}}
\newcommand{\be}{\begin{equation}}
\newcommand{\ee}{\end{equation}}
\newcommand{\bea}{\begin{eqnarray}}
\newcommand{\eea}{\end{eqnarray}}
\newcommand{\beaa}{\begin{eqnarray*}}
\newcommand{\eeaa}{\end{eqnarray*}}

\def\dbE{\mathbb{E}}
\def\dbF{\mathbb{F}}

\def\dbP{\mathbb{P}}
\def\dbR{\mathbb{R}}

\def\dbV{\mathbb{V}}

\def\a{\alpha}

\def\d{\delta}
\def\e{\varepsilon}

\def\k{\kappa}
\def\l{\lambda}

\def\th{\theta}

\def\O{\Omega}
\def\cA{{\cal A}}

\def\cE{{\cal E}}
\def\cF{{\cal F}}

\def\cR{{\cal R}}

\def\no{\noindent}

\def\ms{\medskip}

\def\q{\quad}
\def\qq{\qquad}

\def\cd{\cdot}
\def\cds{\cdots}

\def\qed{ \hfill \vrule width.25cm height.25cm depth0cm\smallskip}

\newcommand{\basa}{\begin{assumption}}
\newcommand{\easa}{\end{assumption}}

\newcommand{\bas}{\begin{assum}}
\newcommand{\eas}{\end{assum}}

\def\limsup{\mathop{\overline{\rm lim}}}

 \def\cd{\cdot}
\def\cds{\cdots}

\def\dis{\displaystyle}

\def\wh{\widehat}

\def\bx{{\bf x}}

\def\1{{\bf 1}}

\def\:{\!:\!}
\def\reff#1{{\rm(\ref{#1})}}
\def \proof{{\noindent \bf Proof\quad}}

at 9pt

\newtheorem{thm}{Theorem}[section]

\newtheorem{rem}[thm]{Remark}
\newtheorem{eg}[thm]{Example}

\newtheorem{assum}[thm]{Assumption}

\begin{document}

\title{ Instability and Efficiency of Non-cooperative Games}
\author{Jianfeng {\sc Zhang}\footnote{Department of Mathematics, University of Southern California, Los
Angeles, CA 90089. E-mail: jianfenz@usc.edu. This author is supported in part by NSF grant DMS-2205972. The author would like to thank  Ali Lazrak and James Enoun  for very helpful comments.}
}\maketitle

\begin{abstract}
It is well known that a non-cooperative game may have multiple equilibria. In this paper we consider the efficiency of games, measured by the ratio between the aggregate payoff over all Nash equilibria and that over all admissible controls. Such efficiency operator is typically unstable with respect to small perturbation of the game. This seemingly bad property can actually be a good news in practice: it is possible that a small change of the game mechanism may improve the efficiency of the game dramatically. We shall introduce a game mediator with limited resources and investigate two mechanism designs aiming to improve the efficiency. Moreover, we compare the mediator's capability of efficiency improvement when she has access to full information or only partial information. When the mechanisms contain only rewards, the mediator has the same power in the two cases. However, when the mediator can use punishments as well, in general she may have a larger power to improve the efficiency in the full information case than in the partial information case. 
\end{abstract}
\vspace{5mm}


\vspace{5mm}

\maketitle

\vfill\eject
\section{Introduction}\label{sec:intro}
\setcounter{equation}{0}
Consider the following two player nonzero-sum game  as an illustrative example:
\begin{table}[h]
  \begin{center}
    \begin{tabular}{|l|c|c|c|} 
     \hline
      $J(a)$ & $a_2 = 0$ & $a_2 = 1$ & $a_2=2$\\
      \hline
      $a_1=0$ & $(100,100)$ & $(0,102)$ & $(0,102)$ \\
      \hline
      $a_1=1$ & $(102,0)$ & $(1,2)$ & $(0,0)$\\
      \hline
      $a_1=2$ & $(102,0)$ & $(0,0)$ & $(3,1)$\\
      \hline
    \end{tabular}
     \caption{\label{tab:illustrative} The illustrative example}
  \end{center}
\end{table}

\vspace{-4mm}
\no where each player $i$ aims to maximize his payoff $J_i(a)$ by choosing his control $a_i$. Clearly this game has two Nash equilibria $a=(1, 1)$ and $(2,2)$, with corresponding payoffs $J(a)=(1,2)$ and $(3,1)$, respectively. It is well known that a game is typically inefficient, in this case both equilibria are much worse in terms of the average payoff (or total payoff) than the socially optimal control $a=(0,0)$ with corresponding payoff $J(a)=(100, 100)$. In fact, here the two equilibria are Pareto dominated by $(0, 0)$, as in the {\it prisoners' dilemma}. In the literature, such inefficiency is measured either by the {\it price of anarchy} corresponding to the worst equilibrium $(1,1)$, see e.g. Koutsoupias-Papadimitriou \cite{KP}, or by the {\it price of stability} corresponding to the best equilibrium $(2,2)$, see e.g. Anshelevich-Dasgupta-et al \cite{ADKTWR}. A natural and important question in the game theory is: 
\bea
\label{question}
\mbox{\it Can we improve the efficiency of a game?}
\eea
In this paper we shall focus on the price of stability\footnote{Both \cite{KP} and \cite{ADKTWR} consider costs, and thus the corresponding prices are greater than $1$. Here we consider payoffs, and hence the "price of stability" or more precisely the efficiency is less than $1$.}, namely on the best equilibrium. We remark that the best equilibrium is automatically Pareto optimal among all equilibria, so at least some players will be happy to implement it. Moreover,   in societies where people tend to trust the authority (the government, or even just the media), once the authority recommends the best equilibrium which is good for the society, each individual player may feel the others (at least most others) will follow and then it is also for his own best interest to follow that equilibrium. That is, it is relatively easier to implement the best equilibrium than to implement the social optimum, from which each player has an incentive to deviate.

One simple answer to the question \reff{question} is the bounded rationality, see e.g. Magill-Quinzii \cite{MQ} and Papadimitriou-Yannakakis \cite{PY}. That is, the players are satisfied with a good enough decision which is not necessarily the best one, then $\e$-equilibria are stable for appropriate $\e$. For example, in Table \ref{tab:illustrative}, if the players are satisfied with $1$-equilibria (namely with $\e=1$),  then $a=(0, 1)$ and $(2, 0)$ become acceptable approximate equilibria, which effectively increases the average payoff from ${1\over 2}[1+3]=2$ to ${1\over 2}[0+102] = 51$. Furthermore, if the players are willing to accept $2$-equilibrium (namely with $\e=2$), then $a=(0,0)$ becomes an acceptable one and thus the game reaches the social optimum with average payoff ${1\over 2}[100+100]=100$. However, in reality this is often not the case, either due to selfishness or due to distrusts among the players, and people indeed get stuck in bad equilibria (thinking of all the internal fights within a society, or even wars among countries). So our goal is to introduce appropriate mechanism designs to improve the efficiency of the game. 

One seminal paper in this direction is Monderer-Tennenholtz \cite{MT1} which proposed the $\k$-implementation.\footnote{\cite{MT1} used the notation $K$, we change to $\k$ since we consider small value here.} See also Bachrach-Elkind-ect al \cite{BEMPZRR}, Deng-Tang-Zheng \cite{DTZ}, Huang-Wang-Wei-Zhang \cite{HWWZ}, Monderer-Tennenholtz \cite{MT2}, Zhang-Farina-et al \cite{ZFACMHCGCS} for some works along this line. In this approach there is a third party called mediator, who can make credible non-negative payments to the players, with the total amount limited to \$$\k$. However, the mediator cannot  design a new game or enforce players’ behavior. She can only provide additional incentives to the players through her payments. The problem is to design the payments in a way such that they would induce the players to implement a desirable outcome, for example the social optimum. 

In this paper we study the problem in an abstract framework, with the $\k$-implementation as one of the two main mechanisms we consider. Instead of targeting at a desirable outcome as in \cite{MT1} which may require a large $\k$, we focus on the effect of a small $\k$, reflecting the reality that quite often the mediator has only limited resources. It is well known that equilibria are very sensitive to small perturbations of the game parameters, and hence the efficiency of the game is overall unstable. This seemingly bad property turns out to be a good news for our purpose: it is possible that a small investment $\k$ could increase the efficiency of the game dramatically.    For example, in Table \ref{tab:illustrative}, when $\k=1$, the mediator can pay  \$1  to Player 1 when the control is $(0,1)$,\footnote{Here we assume the mediator's payments can depend on the players' controls. We shall discuss more on this later.} this will make $(0,1)$ a new equilibrium and hence improve the efficiency significantly. 

By considering the efficiency of the $\k$-game as a function of $\k$, our main result is that this efficiency function is non-decreasing and right continuous in $\k$, but in general it can be discontinuous. The right continuity at $\k=0$ implies that, in order to improve the efficiency of the game, an appropriate amount of investment is needed.\footnote{\cite{MT1} emphasized that it is possible that no  monetary offer is materialized when the players follow the desired behavior. This does not contradict with our result here. In \cite{MT1}, the targeted outcome could already be the best equilibrium, the promised offers just aim to induce the players to implement that one (instead of other equilibria), which do not improve the efficiency. We should also note that, even if the real payment (along the targeted desired outcome) is $0$, the promised offers on the other outcomes may not be small, namely $\k$ could be large. Since the reliability of the mediator is crucial, she has to have \$$\k$ available when making the promises, even if she knows she doesn't need to pay it in the end. Then a large $\k$ will restrict the mediator's ability seriously.} However, this amount is not necessarily large.  It is important to understand this efficiency function in practice, especially the discontinuous points are critical. Indeed, if $\k$ is already close to a discontinuous point, it will be wise to add a little more investment because that small amount of extra investment could increase the efficiency significantly. Moreover, considering the case that the mediator has several projects to take care, each involving a game. Since she may have only limited resources, it will be important to understand which project(s) she should invest her resources. The conventional wisdom would suggest to set a higher priority on the project(s) most important for the society. However, it is possible that a small amount of extra investment makes little or no difference at all to that project(s). Our findings suggest instead that it could be wise to set a higher priority on the project(s) whose efficiency can be increased dramatically by a small investment. We would also like to note that, in the case that the mediator is the government and the payoffs of the base game are individuals' incomes, the government can charge tax on the incomes, then the improvement of the efficiency implies the government will receive more tax. When this extra tax exceeds $\k$ (again noting that $\k$ is small), it could be possible that all players as well as the government increase their gains, and thus the $\k$-implementation will create a win-win-win situation. 

The second main mechanism we investigate is taxation,\footnote{In the previous paragraph, the tax is received afterwards and thus is not considered in the mechanism design. Here, we shall consider the tax as the main resource of the mediator.} again by considering the government as the mediator.  There are numerous publications on tax policies, see e.g. the survey paper Mankiw-Weinzierl-Yagan \cite{MWY} and the references therein. Our focus here is not on optimal tax policy or its impact on the society, but rather on how to redistribute the tax the government receives so as to improve the efficiency of the game. For simplicity, we assume all players are levied with a fixed tax rate $\th\in [0, 1]$ and again we consider only small $\th$.\footnote{Alternatively, we may interpret $\th$ as the incremental increase of the tax rate when the government considers to increase the tax rate, which is typically small. In this paper we assume $\th$ is positive. We may also consider to lower the interest rate and then $\th$ becomes negative, which is covered by our abstract framework but is not investigated  specifically in this paper.}  
This mechanism differs from the $\k$-implementation in two aspects. First, unlike in the $\k$-implementation where each player receives non-negative payments from the mediator, here by first paying tax and then receiving certain redistribution payments, the players could end up receiving punishment, see e.g. Ramirez-Kolumbus-Nagel-Wolpert-Jost \cite{RKNWJ} for mechanism design with both rewarding and punishment. This gives the mediator more power to influence the game. Next, the resource available to the mediator, namely the amount of the tax she receives, depends on the game outcome, while in $\k$-implementation the total resource $\k$ is a fixed constant. Our main result remains true for this mechanism, that is, the efficiency function is non-decreasing and right continuous in $\th$, and may typically be discontinuous in $\th$.

One consequence of the possible punishment is the difference between mechanisms with full information and those with partial information  (or say, with hidden action, as in contract theory). We call a mechanism partially informed if the rewards/punishments depend only on the original payoffs of the game, and fully informed if they can depend on the controls as well. For taxation purpose, it could be more appropriate to use partial information mechanisms since people are required to report their income, but not necessarily their activities. When there are only rewards, we show that the two mechanisms lead to the same effect of efficiency improvement. However, when there is punishment as well, in general full information mechanisms  have  larger power to improve the efficiency than the corresponding  partial information mechanisms. It should be noted though, for the specific taxation mechanisms considered in this paper, since the "punishment" (the tax) is fixed, the mediator can only leverage on the "reward" (the redistribution part), so such difference does not exist. If the mediator can design the tax rate for each player, namely she can design the punishment, then she may indeed gain the power of efficiency improvement in the full information case.

To study the efficiency rigorously, we invoke the set value of the game proposed by Feinstein-Rudloff-Zhang \cite{FRZ}, 
  which roughly speaking is the set of values over all possible equilibria. It is showed in \cite{FRZ} that these set values enjoy stability/regularity in certain sense, which is consistent with our result that the efficiency function is right continuous at $\k=0$ and $\th=0$. The point of this paper is that, such stability/regularity is not uniform in the sense that discontinuity may appear for small $\k$ and $\th$, and thus it  becomes possible and crucial to design mechanisms so as to improve the efficiency of the game.  
 
We shall also discuss briefly two related issues. First, by nature the mechanism design is a leader follower problem, with the mediator as the leader and the players as multiple followers. So the problem is also intrinsically connected to the principal agent problems with multiple agents, see e.g. Segal \cite{Segal}. Next, by reinterpreting the $\th$ in the taxation mechanism as a portion the government will control the economic outcome, then the $\th$ is typically large in a central planned economy.  In the extreme case that $\th=1$, the game problem reduces to a centralized control problem. In this case the payoffs of the base game may already depend on $\th$. We shall investigate the optimization of $\th$, combined with the mechanism design for every given $\th$.
  
Finally we remark that in this paper we only investigate the efficiency of the game in a theoretical way. There are extensive studies on algorithms and their complexity analysis for desired equilibria, which we do not consider in this paper.\footnote{In particular, our efficiency of the game is completely different from the efficiency of numerical algorithms.}  

The rest of the paper is organized as follows. In Section \ref{sect-example} we illustrate our ideas through the example in Table \ref{tab:illustrative}. In Section \ref{sect-full} we study the mechanism schemes with full information, and in Section \ref{sect-partial} we study the partial information case. In Section \ref{sect-weighted} we extend the mediator's problem to weighted average. Section \ref{sect-dynamic} is devoted to continuous time stochastic differential games. We provide further discussions in Section \ref{sect-further} and concluding remarks in Section \ref{sect-conclude}. Finally, in Appendix we complete a technical proof.

\section{The illustrative example}
\label{sect-example}
\setcounter{equation}{0}
Consider the game specified in Table \ref{tab:illustrative}. Denote by $A= A_1\times A_2 =\{0,1,2\}^2$ the admissible control set, and  $J(a) := (J_1(a), J_2(a))$. Introduce the average payoff:
\bea
\label{barJ}
\ol J(a) := {1\over 2}[J_1(a) + J_2(a)].
\eea
This game has two Nash equilibria $\cE = \big\{ (1,1), (2,2)\big\}$.  We define the optimal value of the game problem and the optimal value of the control problem as follows: 
\bea
\label{optimalV}
\left.\ba{c}
\dis V := \sup_{a^*\in \cE} \ol J(a^*)  = \max\Big\{{1+2\over 2}, {3+1\over 2}\Big\} =2;\\
\dis \wh V := \sup_{a\in A} \ol J(a)  = \ol J(0,0) = {1\over 2} [ 100 + 100]=100.
\ea\right.
\eea
We then define the efficiency of the game as\footnote{The efficiency $E$ corresponds to the price of stability in \cite{ADKTWR}. However, \cite{ADKTWR} considers costs, and thus the price of stability is greater than $1$. Here we consider payoffs, and thus our $E$ is less than $1$.}:
\bea
\label{E}
E := {V\over \wh V} = {2\over 100} = 2\%.
\eea
That is, by restricting to equilibria, the players can achieve at most $2\%$ of the optimal value the system could provide them. This is of course a huge waste of the resources, both for the individual players and for the society.\footnote{The fundamental reason for the loss is there are negative externalities embedded in Table \ref{tab:illustrative}. When playing a given strategy, players can impose on the other players huge losses.}  The main goal of this paper is to improve the efficiency by modifying the game slightly. In particular, we shall introduce two possible mechanisms: the $\k$-implementation and the taxation mechanism. 

\begin{rem}
\label{rem-implementation}
(i) We remark that $V$ is a lot easier to achieve than $\wh V$ in practice.  To achieve $\wh V$, one needs to implement $(0,0)$. Since it is not an equilibrium, both players have the incentive to move away from it. To achieve $V$, one needs to implement $(2,2)$, which is an equilibrium. The mediator can simply announce that this is her preferred equilibrium. Note that by nature this equilibrium is Pareto optimal among all equilibria and thus some players (in this case, Player 1) are willing to follow it. Then, as long as the players think the others will follow that (especially when there are a large number of players, instead of two players), this equilibrium will be adopted. 

(ii) It should be emphasized that the best equilibrium $(2,2)$ is Pareto optimal only among the equilibria, but not necessarily Pareto optimal among all controls $a$. In fact, in this example obviously the social optimum $(0,0)$ Pareto dominates the best equilibrium $(2,2)$. 
\end{rem}

\subsection{The $\k$-implementation}
\label{sect-kappa}
Assume the mediator has an extra resource \$$\k$ to add into the system. Given the budget $\k$, a distribution function determines the transfer to both  players. It can therefore be modeled as a function $\pi: A\to \dbR^2$ such that
\bea
\label{psi}
\pi_i(a)\ge 0,~ i=1,2;\q \pi_1(a) + \pi_2(a) \le \k.
\eea
 Let $\Pi_\k$ denote the set of all these distribution functions $\pi$. Introduce
\bea
\label{Jpsi}
J^\pi_i(a) := J_i(a) + \pi_i(a).
\eea
Consider the game with payoffs $J^\pi = (J^\pi_1, J^\pi_2)$, and denote by $\cE(\pi)$ the set of all its equilibria. We introduce the efficiency function as: for the $\wh V$ in \reff{optimalV},
\bea
\label{Epsi} 
\dis E(\k) := \sup_{\pi\in \Pi_\k} E(\pi),\q  E(\pi) := {V(\pi)\over \wh V},\q\mbox{where}\q V(\pi):= \sup\big\{ \ol J(a^\pi): ~ a^\pi\in \cE(\pi)\big\}.
\eea
 Clearly $\Pi_{\k_1} \subset \Pi_{\k_2}$ and thus $E(\k_1)\le E(\k_2)$ when $\k_1 \le \k_2$.
 
 \begin{rem}
\label{rem-kredistribution}
(i) We emphasize that, in \reff{Epsi} we use $\ol J(a^\pi)$, instead of $\ol {J^\pi}(a^\pi)$ which would count the amount ${1\over 2}[\pi_1(a) + \pi_2(a)]$ in the calculation. Note that  $\ol J^\pi(a^\pi)$ is what the two players actually receive, while  $\ol J(a^\pi)$ is what they "produce" or contribute to the society. Since our discussion of efficiency focuses  more  on the interest of the society, and what's really crucial is how the redistribution affect the equilibria, so it makes sense to use $\ol J(a^\pi)$.

(ii) Technically it is also more convenient to study $\ol J(a^\pi)$, because then $\pi$ will only affect the equilibria $\cE(\pi)$, not the average payoff directly. In any case, since we are talking about small $\k$, the difference between the two are minor.

(iii) Since $\hat V$ is fixed (and positive), mathematically it is equivalent to  study the optimization problem $\sup_{\pi\in \Pi_\k} V(\pi)$. In fact this is what we will actually do in later sections where we do not require the payoffs to be positive. Here for illustrative purpose, it is more convenient to consider the ratio which is unit free and is always between $0$ and $1$, then it is easier to see the effect of the efficiency improvement.
\end{rem}

  We next analyze $E(\k)$. When $\k<1$, we are not able to change the equilibria, and the equilibria remain to be $\{(1,1), (2,2)\}$, then by \reff{E} we have
 \beaa
 E(\k) = E(0) = 2\%.
 \eeaa
 When $1\le \k <4$, we can make $\cE(\pi) = \{(0,1),  (1,1), (2,2)\}$\footnote{It's a coincidence that $(1,1)$ and $(2,2)$ remain equilibria for the game with payoffs $J^\pi$. This is in general not true for a general $\pi$. For example, assume $\k=2$ and set $\pi(0,1) = (2, 0)$ and $\pi(a)=0$ otherwise, we see that $(1,1)$ is not an equilibrium anymore.} by setting:
 \beaa
 \pi(0,1) = (1,0), \q  \pi(a)=(0,0) ~\mbox{for all other $a$}.
 \eeaa
 Then
 \beaa
 V(\pi) = \ol J(0,1) = {1\over 2}[0+102] = 51,\q E(\k) = {51\over 100} = 51\% .
 \eeaa
 When $ \k \ge 4$, we can make $\cE(\pi) = \{(0,0),  (1,1), (2,2)\}$ by setting:
 \bea
 \label{k=4}
 \pi(0,0) = (2,2), \q  \pi(a)=(0,0) ~\mbox{for all other $a$}.
 \eea
 Then
 \beaa
 V(\pi) = \ol J(0,0)= {1\over 2}[100+100] = 100,\q E(\k) = {100\over 100} = 100\% .
 \eeaa
 In summary,
 \bea
\label{Ek2}
E(\k) = \left\{\ba{lll} 2\%,\qq 0\le \k < 1;\\ 51\%,\q~ 1\le \k < 4;\\ 100\%,\q\k\ge 4.\ea\right.
\eea
That is, by investing \$$1$ into the system, the mediator may improve the efficiency of the game from $2\%$ to $51\%$. If she can invest \$$4$, the efficiency can increase further to $100\%$.

\begin{rem}
\label{rem-unstable} 
It is important to understand the efficiency function like \reff{Ek2} in practice.

(i) As we can see, $E(\k)$ is discontinuous in $\k$. It is crucial to figure out these discontinuous points. For example, if the mediator has already invested $\k = 0.99$, it is better to add a little more investment to increase it to $\k = 1$ so that the efficiency of the game can increase from $2\%$ to $51\%$. Similarly, it is not wise to invest $\k=3.99$, one would rather increase to $\k=4$ in that case.

(ii) Consider the situation that the mediator (say a government) takes care of two projects, each involving a game. The mediator has some limited resources and can invest only in one project. The conventional wisdom would suggest to invest in the project more important for the society. However, our analysis shows that it is possible that such a small investment may make little (or no) impact on that project but can improve the efficiency (or total welfare) of the other project significantly. In that scenario, it could be wiser to invest in the other project even though it is less important. 
 \end{rem}
 
\begin{rem}
\label{rem-tax}
Now assume the mediator is the government (and nevertheless still consider two players for illustrative purpose), and it can levy income tax afterwards with rate $\th$. For simplicity assume the payment $\pi$ the players receive from the government is tax free. So for the original base game, the best equilibrium is $(2,2)$, then Player 1, Player 2, and the government will receive, respectively:
\beaa
P1:  3(1-\th),\q P2: 1-\th,\q G: 4\th.
\eeaa 
When $\k=4$,  with the $\pi$ in \reff{k=4}, the best equilibrium is $(0,0)$, the payoffs they will receive become  (after the transfer payoffs from the government): 
\beaa
P1: 2 + 100(1-\th),\q P2: 2+100(1-\th),\q G: 200\th - 4.
\eeaa
Let's say $\th = 5\%$, then both players as well as the government will have a larger payoff:
\begin{table}[h]
  \begin{center}
    \begin{tabular}{|l|c|c|c|c|} 
     \hline
       & P1 & P2 & G & E\\
      \hline
      $\k=0$ & $2.85$ & $0.95$ & $0.2$ & $2\%$ \\
         \hline
      $\k=4$ & $97$ & $97$ & $6$ & $100\%$\\
      \hline
    \end{tabular}
     \caption{\label{tab:illustrative-comparison}  win-win-win example}
  \end{center}
\end{table}

\no So, by improving the efficiency, this is a win-win-win situation.
\end{rem}

\subsection{A taxation mechanism}
\label{sect-theta}
Inspired by Remark \ref{rem-tax}, actually the government doesn't have to invest extra resources. It can just invest the tax it's going to receive (after all the government does not make money by itself).\footnote{Instead of the actual tax it receives, we may also view the right of making tax policy as the resource of the government. Then a small $\th$ amounts to a limited resource.} That is, assume the tax rate is $\th\in [0,1]$, the government can introduce a distribution function $\pi: A\to \dbR^2$ such that
\bea
\label{psi2}
\pi_i(a)\ge 0,~ i=1,2;\q \pi_1(a) + \pi_2(a) \le \th [J_1(a) + J_2(a)].
\eea
Let $\Pi_\th$ denote the set of all these functions $\pi$. By abusing the notation with \reff{Jpsi}, introduce
\bea
\label{Jpsi2}
J^\pi_i(a) := (1-\th)J_i(a) + \pi_i(a).
\eea
Then, for the $E(\pi)$ defined in \reff{Epsi} corresponding to this $J^\pi$,   we define
\bea
\label{Eth}
E(\th) := \sup_{\pi \in \Pi_\th}E(\pi).
\eea

\begin{rem}
\label{rem-mredistribution}
(i) We should note that in the previous subsection $\ol {J^\pi}(a^\pi)\ge \ol J(a^\pi)$, while here $\ol {J^\pi}(a^\pi)\le \ol J(a^\pi)$. Indeed, players pay taxes and receive transfers. When the transfers are less than the taxes they pay, they are net payers. 
In particular,  this setting includes the mechanism of punishment. The government can discourage the players to choose some controls, e.g. $(1,1)$ and $(2,2)$, through punishment by setting $\pi(1,1) = \pi(2,2) = (0,0)$. 

(ii) This setting also includes the mechanism of rewarding. For example, if we set $\pi(0,1) = (102\th, 0)$, then $J^\pi(0, 1) = (102\th, 102(1-\th)$ and thus we are rewarding Player 1 in this case. Moreover, since the tax is levied proportionally, so the rich players are contributing more.  
\end{rem}

\begin{rem}
\label{rem-boundedrationality}
(i) A special case of  the taxation mechanism is the uniform redistribution: $\pi_i(a) := {\th\over 2}[J_1(a)+J_2(a)]$. Then  \reff{Jpsi2} becomes:
\bea
\label{Jpsi3}
J^\pi_i(a) := (1-\th)J_i(a) + \th \ol J(a).
\eea

(ii) Alternatively, one may interpret the model \reff{Jpsi3} with the bounded rationality (cf. \cite{MQ}). That is, the individual players are willing to sacrifice his/her own utility for the benefit of the whole society to certain degree, and $\th$ is a measure for this degree. 

(iii) The model \reff{Jpsi3} provides a natural bridge between the decentralized game problem (when $\th=0$) and the centralized control problem (when $\th=1$). So $\th$ can also be viewed as a degree of centralization. 
\end{rem}

In the spirit of Remark \ref{rem-mredistribution}, given $\th$, we would like to choose $\pi$ to encourage "good" outcomes and discourage "bad" outcomes. By direct analysis, the best $\pi^*$ and then the corresponding $J^{\pi^*}$ should be as in Table \ref{tab:illustrative3}. Consequently, the efficiency function is as in Table \ref{tab:illustrative4}. In summary,
\bea
\label{Eth1}
E(\th)  = E(\pi^*)= \left\{\ba{lll} 2\%,\qq 0\le \th < {1\over 103};\\ 51\%,\q~ {1\over 103}\le \th < {1\over 51};\\ 100\%,\q {1\over 51} \le \th \le 1.\ea\right.
\eea
\begin{table}[h]
  \begin{center}
    \begin{tabular}{|c|c|c|c|} 
     \hline
      $\pi^*$ & $a_2 = 0$ & $a_2 = 1$ & $a_2=2$\\
      \hline
      $a_1=0$ & $(100\th,100\th)$ & $(102\th,0)$ & $(102\th,0)$ \\
      \hline
      $a_1=1$ & $(0, 102\th)$ & $(0,0)$ & $(0,0)$\\
      \hline
      $a_1=2$ & $(0, 102\th)$ & $(0,0)$ & $(0,0)$\\
      \hline
    \end{tabular}\qq\q
      \begin{tabular}{|l|c|c|c|} 
     \hline
      $J^{\pi^*}(a)$ & $a_2 = 0$ & $a_2 = 1$ & $a_2=2$\\
      \hline
      $a_1=0$ & $(100,100)$ & $(102\th,102(1-\th))$ & $(102\th,102(1-\th))$ \\
      \hline
      $a_1=1$ & $(102(1-\th),102\th)$ & $(1-\th,2(1-\th))$ & $(0,0)$\\
      \hline
      $a_1=2$ & $(102(1-\th),102\th)$ & $(0,0)$ & $(3(1-\th),1-\th)$\\
      \hline
    \end{tabular}
     \caption{\label{tab:illustrative3} The  taxation mechanism}
  \end{center}
\end{table}
\begin{table}[h]
  \begin{center}
      \begin{tabular}{|c|c|c|c|} 
     \hline
      $\th$ & $\cE(\pi^*)$ & best NE & $E(\th) = E(\pi^*)$\\
      \hline
      $\th< {1\over 103} $ & $(1,1), (2,2)$ & $(2,2)$ & $2\%$ \\
      \hline
      $\th= {1\over 103}$ & $(0,1), (1,0), (1,1), (2,2)$ & $(0,1), (1,0)$ & $51\%$\\
      \hline
      ${1\over 103} < \th < {1\over 51}$ & $(0,1),  (1,0)$ & $(0,1),  (1,0)$ & $51\%$\\
       \hline
      $\th ={1\over 51}$ & $(0,0), (0,1), (1,0)$ & $(0,0)$ & $100\%$\\
      \hline
      ${1\over 51} < \th \le 1$ & $(0,0)$ & $(0,0)$ & $100\%$\\
      \hline
    \end{tabular}
     \caption{\label{tab:illustrative4} The efficiency in the taxation mechanism}
  \end{center}
\end{table}

\no That is, by levying ${1\over 103} \approx 1\%$ of tax and by designing the redistribution mechanism $\pi$ optimally, one may improve the efficiency of the game from $2\%$ to $51\%$. If one can levy ${1\over 51} \approx 2\%$ of tax, the efficiency can increase to $100\%$. Moreover, as in Remark \ref{rem-unstable}, we see that $E(\th)$ is discontinuous in $\th$, and it is crucial to figure out these discontinuous points, which are ${1\over 103}$ and ${1\over 51}$ in this example. It's not wise to set a tax rate right below these discontinuous points.

\section{Mechanism schemes with full information}
\label{sect-full}
\setcounter{equation}{0}
In this section we consider a general static $N$-player game. Player $i$ has control set $A_i$, and  denote $A := A_1\times \cds\times A_N$. Given a control $a =(a_1,\cds, a_N)\in A$, player $i$ will receive payoff $J_i(a)$, which can be negative in general, and denote $J(a):= (J_1(a),\cds, J_N(a))$.  We say $a^*\in A$ is an equilibrium for the game with payoff $J$, denoted as $a^*\in \cE$, if 
\bea
\label{NE}
J_i(a^*) \ge J_i(a^{*,-i}, a_i),\q \forall a_i\in A_i, \forall i=1,\cds, N.
\eea
From now on, we assume $\cE \neq \emptyset$.\footnote{\label{chaos}When $\cE = \emptyset$, we may define $V := -\infty$ in \reff{V}. The interpretation is that, in this case the system could be in chaos, and in many practical situations, a bad equilibrium is still better than chaos.}
We then define the optimal aggregated game value as:
\bea
\label{V}
 V:= \sup_{a^*\in \cE} \ol J(a^*), \q\mbox{where}\q \ol J(a) := {1\over N}\sum_{i=1}^N J_i(a).
\eea

We next turn to the mechanisms the mediator would take. We say the mediator has full information if she observes the control $a$. In this case, she may choose a distribution function $\pi: A\to \dbR^N$, called {\it  full information mechanism}, and we let $\wh\Pi^f$ denote the set of all these mechanisms. For each $\pi\in \wh\Pi^f$, define $J^\pi$ by \reff{Jpsi}.
Let $\cE_\pi$ denote the set of equilibria for the game with payoffs $J^\pi$, and we introduce the game's optimal value:
\bea
\label{Vpi}
V(\pi) := \sup\{\ol J(a^*): a^*\in \cE_\pi\}.
\eea
We emphasize again that, as in Remark \ref{rem-kredistribution}, here we use $\ol J$, rather that $\ol{J^\pi}$. Moreover, as in Footnote \ref{chaos}, we set $V(\pi):= -\infty$ when $ \cE_\pi = \emptyset$.

We now call any subset $\Pi\subset \wh\Pi^f$  a {\it mechanism scheme}, which consists of all possible mechanisms the mediator could choose. For example, for the $\k$-implementation, we have
\bea
\label{Pifk}
\Pi^f_\k:= \big\{\pi\in \wh\Pi^f: \pi_i(a)\ge 0,~ \sum_{i=1}^N \pi(a) \le \k, ~\forall a\in A\big\}.
\eea
Note that in this section both $J$ and $\pi$ can be negative in general, and then the optimal value of the control problem: $\wh V := \sup_{a\in A} \ol J(a)$, can also be negative.  As pointed out in Remark \ref{rem-kredistribution} (iii), in this case it makes more sense to maximize $V(\pi)$ directly, rather than the efficiency $E(\pi):= {V(\pi)\over \wh V}$. So, given a mechanism scheme $\Pi$, our target problem is:
\bea
\label{VPi}
V(\Pi) := \sup_{\pi\in \Pi} V(\pi).
\eea
However, we shall nevertheless still call $V(\Pi)$ the efficiency of the game.

Introduce the Hausdorff distance for the sets $\Pi$:
\bea
\label{dPi}
\left.\ba{c}
\dis d(\pi, \Pi) := \inf_{\tilde \pi\in \Pi} \|\pi- \tilde \pi\|_\infty, \q \|\pi- \tilde \pi\|_\infty := \sup_{a\in A} |\pi(a)-\tilde \pi(a)|,\q \forall \pi\in \wh\Pi^f,~ \Pi\subset \wh\Pi^f;\\
\dis d(\Pi_1, \Pi_2) := \max\Big(\sup_{\pi_1\in \Pi_1} d(\pi_1, \Pi_2),~\sup_{\pi_2\in \Pi_2} d(\pi_2, \Pi_1)\Big),\q\forall \Pi_1, \Pi_2\subset \wh\Pi^f.
\ea\right.
\eea
Our main result of this section is as follows.
\begin{thm}
\label{thm-mon}
(i) $V$ is increasing in $\Pi$ in the sense that: 
\beaa
V(\Pi_1) \le V(\Pi_2),\q\mbox{whenever}\q \Pi_1 \subset \Pi_2.
\eeaa

(ii) Assume $A$ is finite,\footnote{This condition is just for technical simplicity. By introducing set values for the games, which involve approximate equilibria, the result here holds true in much more general framework. See Section \ref{sect-dynamic}.} then $V$ is upper semi-continuous on compact $\Pi$  in the sense that: when $\Pi$ is compact,  
\beaa
\limsup_{n\to\infty} V(\Pi_n) \le V(\Pi), \q\mbox{whenever}\q \lim_{n\to\infty} d(\Pi_n, \Pi) =0.
\eeaa
 In particular, if $\Pi$ is compact, $\Pi_n \downarrow \Pi$ and $d(\Pi_n, \Pi) \to 0$, then $V(\Pi_n) \downarrow V(\Pi)$.
\end{thm}
\proof By \reff{Vpi}, (i) is obvious. To see (ii), let $\Pi_n\to \Pi$ and $\Pi$ be compact. Fix a sequence $\e_n\to 0$. For each $n\ge 1$, there exist $\pi_n\in \Pi_n$ and $a^*_n \in \cE_{\pi_n}$ such that
\bea
\label{VPin}
V(\Pi_n) \le  V(\pi_n)+\e_n  \le \ol J(a^*_n) +2\e_n. 
\eea
By \reff{dPi}, there exists $\pi_n'\in \Pi$ such that $\|\pi_n' - \pi_n\|_\infty \le d(\Pi_n, \Pi) + \e_n$. Moreover, since $A$ is finite and $\Pi$ is compact, by possibly following a subsequence, we have $a_n \to a^*\in A$ and $\pi'_n \to \pi^*\in \Pi$. Denote $\d_n := d(\Pi_n, \Pi)  \vee |a_n^*-a^*|\vee \|\pi_n'-\pi^*\|_\infty\vee \e_n \to 0$. Then
\beaa
\|\pi_n - \pi^*\|_\infty \le \|\pi_n - \pi_n'\|_\infty + \|\pi_n'-\pi^*\|_\infty \le d(\Pi_n, \Pi) + \e_n + \|\pi_n'-\pi^*\|_\infty  \le 3\d_n.
\eeaa 
Recall \reff{NE}, for any $i$ and $a_i$ we have
\beaa
J^{\pi^*}_i((a^*_n)^{-i}, a_i) &\le& J^{\pi_n}_i((a^*_n)^{-i}, a_i) + \|\pi_n - \pi^*\|_\infty \le J^{\pi_n}_i(a^*_n) + \|\pi_n - \pi^*\|_\infty\\
&\le&J^{\pi^*}_i(a^*_n)  + 2\|\pi_n - \pi^*\|_\infty  \le J^{\pi^*}_i(a^*_n)  + 6\d_n.
\eeaa
Note that $|A|<\infty$, then $a_n = a^*$ for $n$ large enough. Thus, for $n$ large, we have
\beaa
J^{\pi^*}_i((a^*)^{-i}, a_i)  \le J^{\pi^*}_i(a^*)  + 6\d_n.
\eeaa
Send $n\to \infty$, we see that $a^*\in \cE_{\pi^*}$. Therefore, for $n$ large enough, by \reff{VPin} have
\beaa
V(\Pi_n)  \le \ol J(a^*_n) +2\e_n = \ol J(a^*) + 2\e_n \le V(\pi^*) + 2\e_n \le V(\Pi) + 2\e_n. 
\eeaa
Note that we are talking about subsequence. The above implies that, for any subsequence $\{n_k\}$, there exists a further subsequence $n_{k_l}$ such that $\lim_{l\to\infty} V(\Pi_{n_{k_l}}) \le V(\Pi)$.  One can easily see that this  implies $\limsup_{n\to\infty} V(\Pi_n) \le V(\Pi)$.

Finally, when $\Pi_n \downarrow \Pi$, since $V(\Pi_n)\ge V(\Pi)$ and $\limsup_{n\to\infty} V(\Pi_n) \le V(\Pi)$, we obtain $V(\Pi_n) \downarrow V(\Pi)$ immediately.
\qed

 Inspired by Subsections \ref{sect-kappa} and \ref{sect-theta}, we  have the following two examples.
 
 \begin{eg}
 \label{eg-pi}
 (i) Recalling \reff{Pifk}, the mapping $\k\mapsto V(\Pi^f_\k)$ is increasing. Moreover, when $A$ is finite,  clearly $\Pi_\k$ is compact and thus $V(\Pi^f_\k)$ is right continuous in $\k$.
 
 (ii) Let $J_i\ge 0$, and for $\th\in [0,1]$, $\Pi^f_\th$ denote the set of functions $\pi: A\to \dbR^N$ satisfying\footnote{We are abusing the notation here with those in Subsection \ref{sect-theta}. The $\pi$ there corresponds to the $\psi$ here.}
 \bea
 \label{Pifth}
\pi_i(a) = \psi_i(a) - \th J_i(a),\q \mbox{where}\q \psi_i(a) \ge 0, ~i=1,\cds, N,~ \sum_{i=1}^N \psi_i(a) \le \th\sum_{i=1}^N J_i(a).
 \eea
 Then the mapping $\th\mapsto V(\Pi^f_\th)$ is increasing.  Moreover, when $A$ is finite,  $\Pi^f_\th$ is compact and thus $V(\Pi^f_\th)$ is also right continuous in $\th$. We remark that $\pi$ can be negative here.
 \end{eg}

\begin{rem}
\label{rem-stable}
The right continuity in the above example indicates that there is no free lunch: in order to improve the efficiency, the mediator needs to make some efforts. However, such regularity is not uniform, in particular, in Example \ref{eg-pi} $V$ may have discontinuity at small $\k$ or small $\th$. These discontinuous points of $\k$ or $\th$ are crucial for our mechanism design, as pointed out in Remark \ref{rem-unstable}.
\end{rem}

\section{Mechanism schemes with partial information}
\label{sect-partial}
\setcounter{equation}{0}
Consider the setting in the previous section again. In many practical situations, the mediator may observe $J(a)$, but not the players' control $a$. For example, for the taxation scheme with the government as the mediator, people are required to report their income, but not necessarily their activities. In this case, as we did in the previous section, the mediator would  choose a distribution function $\pi: \cR\to \dbR^N$, called {\it partial information mechanism}, where $\cR := \{J(a): a\in A\}$ is the range of all payments.  We let $\wh\Pi^p$ denote the set of all partial information mechanisms, and for each $\pi\in \wh\Pi^p$, define $J^\pi$ as in \reff{Jpsi}:
\bea
\label{Jpi-partial}
J^\pi_i(a) := J_i(a) + \pi_i(J(a)).
\eea

\begin{rem}
\label{rem-information}
For illustrative purpose, in this remark we use $\pi^f$ and $\pi^p$ to denote representative full information and partial information mechanisms, respectively. For the rest of the paper, we shall abuse the notation and denote both by $\pi$. 

(i) In the full information case, the mediator observes $J(a)$ as well, so supposedly she can choose mechanism in the form $\tilde \pi^f(a, J(a))$. However, since $J$ is given, this is equivalent to choose $\pi^f(a):= \tilde \pi^f(a, J(a))$. 

(ii) For any $\pi^p\in \wh \Pi^p$, again since $J$ is known, we may naturally introduce a corresponding $\pi^f\in \wh\Pi^f$: $\pi^f(a) := \pi^p(J(a))$. In this sense, we claim $\wh \Pi^p \subset \wh \Pi^f$.

(iii) For any $\pi^f\in \wh\Pi^f$, there exists a corresponding $\pi^p \in \wh \Pi^p$ in the sense of (ii) above if and only if $\pi^f$ is $J$-invariant, that is, $\pi^f(a) = \pi^f(\tilde a)$ whenever $J(a) = J(\tilde a)$.

Later on, we shall use the same notation $\pi$, and we state (iii) as: a mechanism $\pi \in \wh\Pi^f$ is in $\wh\Pi^p$ if and only if it is $J$-invariant. 
\end{rem} 

\begin{rem}
 \label{rem-psiy}
 It is possible that there exist an equilibrium $a^*\in \cE$ and a control $a\notin\cE$  such that $J(a) = J(a^*)$. For example, in Table \ref{tab:y-a}, $J(0,0) = J(0,1) = (1,1)$, while $a^*=(0,0)$ is an equilibrium, but $a=(0,1)$ is not.
   \begin{table}[h]
  \begin{center}
    \begin{tabular}{|l|c|c|c|} 
     \hline
      $J(a)$ & $a_2 = 0$ & $a_2 = 1$ \\
      \hline
      $a_1=0$ & $(1,1)$ & $(1,1)$  \\
      \hline
      $a_1=1$ & $(0,0)$ & $(2,2)$ \\
      \hline
    \end{tabular}
     \caption{\label{tab:y-a} two controls sharing the same value }
  \end{center}
\end{table}
 \end{rem}

Now an interesting question is:  compared with the full information case,
\bea
\label{question2}
\mbox{\it do partially informed mediators  have less power to improve the efficiency?} 
\eea
In particular, recall \reff{Pifk}, \reff{Pifth}, and introduce 
\bea
\label{Pip}
\left.\ba{lll}
\dis \Pi^p_\k:= \big\{\pi\in \wh\Pi^p: \pi_i(y)\ge 0,~ \sum_{i=1}^N \pi(y) \le \k, ~\forall y\in \cR\big\}.\\
\dis \Pi^p_\th:= \big\{\pi\in \wh\Pi^p: \pi_i(y) = \psi_i(y)- \th y_i,~ \psi_i(y)\ge 0,~ \sum_{i=1}^N \psi_i(y) \le \th\sum_{i=1}^N y_i,~\forall y\in \cR\big\}.
\ea\right.
\eea
The question is: {\it do we have $V(\Pi^f_\k) = V(\Pi^p_\k)$ and $V(\Pi^f_\th) = V(\Pi^p_\th)$?} 

The following result provides a positive answer to this question.  Let $D: \cR \to 2^{\dbR^N}$ be a measurable mapping such that $D(y)\subset \dbR^N$ is a closed set for each $y\in \cR$.  Introduce 
\bea
\label{PiD}
\left.\ba{lll}
\dis \Pi^f_D:= \big\{\pi\in \wh\Pi^p: \pi(a)\in D(J(a)), \forall a\in A\big\}; ~ \Pi^p_D:= \big\{\pi\in \wh\Pi^p: \pi(y) \in D(y),\forall y\in \cR\big\}.
\ea\right.
\eea 
By Remark \ref{rem-information} and Theorem \ref{thm-mon} (i), it is clear that  $\Pi^p_D \subset \Pi^f_D$ and $V(\Pi^p_D)\le V(\Pi^f_D)$.
  \begin{thm}
 \label{thm-partial}
 Assume $D(y)\subset \dbR_+^N$ for each $y\in \dbR^N$. Then $V(\Pi^f_D) = V(\Pi^p_D)$.
 
 In particular, $V(\Pi^f_\k) = V(\Pi^p_\k)$ and, when $J_i\ge 0$,  $V(\Pi^f_\th) = V(\Pi^p_\th)$.
\end{thm}
\proof (i) In this step we prove the general result. It suffices to prove $V(\Pi^f_D)\le V(\Pi^p_D)$.

Fix arbitrary $\pi^f\in  \Pi^f_D$ and $a^*\in \cE_{\pi^f}$. Introduce $\pi^p\in \Pi^p_D$ by: 
\bea
\label{partial-full}
\pi^p(y) := \pi^f(a^*)~\mbox{ for}~ y= J(a^*),\q\mbox{ and}\q \pi^p(y) := 0 ~\mbox{for}~ y\in \cR\backslash \{J(a^*)\}.
\eea
  Now fix arbitrary $i=1,\cds, N$ and $a_i\in A_i$. If $J(a^{*, -i}, a_i)= J(a^*)$, then $\pi^p(J(a^{*, -i}, a_i)) = \pi^p(J(a^*)) = \pi^f(a^*)$,  and thus
\beaa
&&\dis J^{\pi^p}_i(a^{*, -i}, a_i) - J^{\pi^p}_i(a^*) = J(a^{*, -i}, a_i) - J(a^*)  = 0. 
\eeaa
If $J(a^{*, -i}, a_i)\neq J(a^*)$, then $\pi^p_i(J(a^{*, -i}, a_i)) = 0 \le \pi^f_i(a^{*, -i}, a_i)$, where the inequality thanks to the assumption that $\pi^f(a^{*, -i}, a_i)\in D(J(a^{*, -i}, a_i))\subset \dbR_+^N$.  Note further that $\pi^p(J(a^*)) = \pi^f(a^*)$, thus
\beaa
&&\dis J^{\pi^p}_i(a^{*, -i}, a_i) - J^{\pi^p}_i(a^*) \le J^{\pi^f}_i(a^{*, -i}, a_i) - J^{\pi^f}_i(a^*)\le 0. 
\eeaa
This implies that $a^* \in \cE_{\pi^p}$. Since $\pi^f\in  \Pi^f_D$ and $a^*\in \cE_{\pi^f}$ are arbitrary, we obtain the desired inequality $V(\Pi^f_D)\le V(\Pi^p_D)$. 

(ii) For the $\k$-implementation, clearly it corresponds to 
\beaa
D(y) \equiv D_\k:= \big\{\tilde y\in \dbR^N: y_i\ge 0, \sum_{i=1}^N y_i \le \k\big\} \subset \dbR_+^N,
\eeaa
and thus by (i) we have $V(\Pi^f_\k) = V(\Pi^p_\k)$.

For the taxation mechanism, note that $\pi$ can be negative. However, we may reformulate the game to an equivalent one. Recall the structure $\pi^f_i(a) = \psi^f_i(a) - \th J_i(a)$ and $\pi^p_i(y) = \psi^p_i(y) - \th y_i$. Introduce
\beaa
\tilde J^{\psi^f}(a) := J(a) + {1\over 1-\th} \psi^f_i(a),\q  \tilde J^{\psi^p}(a) := J(a) + {1\over 1-\th} \psi^p_i(J(a)).
\eeaa
Then clearly $J^{\pi^f}(a) = (1-\th) \tilde J^{\psi^f}(a)$ and $J^{\pi^p}(a) = (1-\th) \tilde J^{\psi^p}(a)$. Since $1-\th >0$, then $\cE_{\pi^f} = \cE_{\psi^f}$ and $\cE_{\pi^p} = \cE_{\psi^p}$. Note that, by viewing $\psi$ as the mechanism, we have
\beaa
D(y) := D_\th(y) := \big\{\tilde y\in \dbR^N: \tilde y_i \ge 0, \sum_{i=1}^N \tilde y_i \le \th \sum_{i=1}^N y_i\big\}\subset \dbR_+^N.
\eeaa
We thus obtain $V(\Pi^f_\th) = V(\Pi^p_\th)$ from (i) again.
\qed

\begin{rem}
\label{rem-partialpositive}
(i) The condition $\pi \in D \subset \dbR_+^N $ means the mechanisms has rewarding only. In this case, the full information and the partial information would have the same efficiency improving effect. For the taxation mechanism, although $\pi$ can be negative, however, the punishment part $\th J_i(a)$ is fixed, and the mediator can only leverage the redistribution part $\psi$, which is again nonnegative. 

(ii) When the mediator can indeed design the punishment mechanism, then the mediator could have larger power to improve the efficiency of the game in the full information case than the partial information case, as we see in the following example. 
\end{rem}

 \begin{eg}
 \label{eg-partial}
 For the following $D$, in general $V(\Pi^p_D) < V(\Pi^f_D)$: 
  \bea
 \label{pinegative}
D(y)  := \big\{\tilde y\in \dbR^N: \tilde y_i \le 0, ~ \sum_{i=1}^N \tilde y_i \ge -1\big\},\q \forall y.
 \eea
 \end{eg}
 \proof Consider the following example in Table \ref{tab:open}. We  construct an $\pi^f\in \Pi^f_D$, and both $\pi^f$ and $J^{\pi^f}$ are also reported in Table \ref{tab:open}.

 \begin{table}[h]
  \begin{center}
    \begin{tabular}{|l|c|c|c|c|} 
     \hline
      $J(a)$ & $a_2 = 0$ & $a_2 = 1$ & $a_2=2$ \\
      \hline
      $a_1=0$ & $(100,100)$ & $(101,101)$ & $(1,1)$ \\
      \hline
      $a_1=1$ & $(101,101)$ & $(1,1)$ & $(2,103)$ \\
      \hline
      $a_1=2$ & $(1,1)$ & $(103,2)$ & $(1,1)$ \\
      \hline
    \end{tabular}\ms\\
    \begin{tabular}{|l|c|c|c|c|} 
     \hline
      $\pi^f(a)$ & $a_2 = 0$ & $a_2 = 1$ & $a_2=2$ \\
      \hline
      $a_1=0$ & $(0,0)$ & $(0,-1)$ & $(0,0)$ \\
      \hline
      $a_1=1$ & $(-1,0)$ & $(0,0)$ & $(0,0)$ \\
      \hline
      $a_1=2$ & $(0,0)$ & $(0,0)$ & $(0,0)$ \\
      \hline
    \end{tabular}\q \begin{tabular}{|l|c|c|c|c|} 
     \hline
      $J^{\pi^f}(a)$ & $a_2 = 0$ & $a_2 = 1$ & $a_2=2$ \\
      \hline
      $a_1=0$ & $(100,100)$ & $(101,100)$ & $(1,1)$ \\
      \hline
      $a_1=1$ & $(100,101)$ & $(1,1)$ & $(1,103)$ \\
      \hline
      $a_1=2$ & $(1,1)$ & $(103,1)$ & $(1,1)$ \\
      \hline
    \end{tabular}
     \caption{\label{tab:open} a full information example}
  \end{center}
\end{table}
  One can verify straightforwardly that $(0,0)\in\cE_{\pi^f}$. Then
  \bea
  \label{VPif}
  V(\Pi^f_D) \ge V(\pi^f) \ge \ol J(0,0) = {1\over 2}[100+100] = 100.
  \eea
On the other hand, for any $\pi^p\in\Pi^p_D$, we claim that
\bea
\label{partial-claim}
(0,0), (0,1), (1,0) \notin \cE_{\pi^p},\q (1,2), (2,1)\in \cE_{\pi^p}.
\eea
Then it is clear that $V(\pi^p) = {1\over 2}[1+103] = 52$. Since $\pi^p\in\Pi^p_D$ is arbitrary, we have 
\beaa
V(\Pi^p_D) = 52 < 100 \le V(\Pi^f_D).
\eeaa

 We first verify \reff{partial-claim} for $a=(0,0)$. Since $J(0,1)= J(1,0) = (101, 101)$, by the nature of partial information mechanisms, we have $J^{\pi^p}(0,1) = J^{\pi^p}(1,0)$. Note that
\beaa
J^{\pi^p}_1(0,1) + J^{\pi^p}_2(0,1) = 202 + \pi_1(101,101) + \pi_2(101,101) \ge 201.
\eeaa
We must have $J^{\pi^p}_2(0,1) \ge 100.5$ or $J^{\pi^p}_1(1,0) = J^{\pi^p}_1(0,1) \ge 100.5$. However, since $\pi^p_i\le 0$, we have  $J^{\pi^p}_1(0,0) \le 100, J^{\pi^p}_2(0,0) \le 100$. Then either $J^{\pi^p}_1(0,0) <  J^{\pi^p}_1(1,0)$ or $J^{\pi^p}_2(0,0) <  J^{\pi^p}_2(0,1)$, thus $(0,0)\notin \cE_{\pi^p}$. 

Next, for $a=(0,1)$,  by \reff{pinegative} we have $-1 \le \pi^p_i \le 0$, then 
\beaa
J^{\pi^p}_1(0,1)\le  101 < 102 \le  J^{\pi^p}_1(2,1).
\eeaa
This implies that $(0,1)\notin \cE_{\pi^p}$. Similarly, $(1,0)\notin \cE_{\pi^p}$.

Finally, for $a=(1,2)$, again since $-1\le \pi^p_i\le 0$, we have
\beaa
&J^{\pi^p}_2(1,0) \le 101 < 102 \le J^{\pi^p}_2(1,2),\q  J^{\pi^p}_2(1,1) \le 1 < 102 \le J^{\pi^p}_2(1,2);\\
&J^{\pi^p}_1(0,2) \le 1  \le J^{\pi^p}_1(1,2),\q  J^{\pi^p}_1(2,2) \le 1 \le J^{\pi^p}_1(1,2).
\eeaa
This implies that $(1,2)\in \cE_{\pi^p}$. Similarly we have $(2,1)\in \cE_{\pi^p}$.
\qed

\section{Weighted efficiency of games}
\label{sect-weighted}
\setcounter{equation}{0}
In \reff{V} and \reff{Vpi}, we considered the average value $\ol J$ with equal weights when choosing the best equilibria. In practice, however, some players are more important for the society than some others.\footnote{Instead of individual players, we may also think of some sectors which are critical for the society.} So, when choosing the best equilibria, we may want to set a higher priority on those players. This can be achieved by considering a weighted average:
\bea
\label{weightolJ}
\ol J_\l(a) := \sum_{i=1}^N \l_i J_i(a),\q\mbox{where}\q \l_i>0,~ \sum_{i=1}^N \l_i = 1.
\eea
Clearly, a larger $\l_i$ implies the higher priority of the player $i$, and the $\ol J$ in the previous sections corresponds to the case $\l_1=\cds=\l_N={1\over N}$. We then similarly define
\bea
\label{weightVpi}
V_\l := \sup\big\{\ol J_\l(a^*): a^*\in \cE\big\},~ V_\l(\pi) := \sup\big\{\ol J_\l(a^*): a^*\in \cE_\pi\big\},~ V_\l(\Pi) := \sup_{\pi\in \Pi} V_\l(\pi).
\eea
For fixed $\l$, all the results in the previous two sections remain true, under obvious modifications. 

\begin{rem}
\label{rem-setvalue2}
Since $\l_i>0$, as in Remark \ref{rem-implementation}, the best equilibrium here is Pareto optimal among all equilibria, but may not be Pareto optimal among all admissible controls.
\end{rem}

\begin{rem}
\label{rem-weight}
(i) The weights $\l$ are only under the consideration of the mediator. It should be emphasized that the individual players do not care about $\l$, so the $\l$ does not change the game structure. In particular, the equilibria $\cE$ and $\cE_\pi$ do not depend on $\l$, but the aggregated optimal game values $V_\l, V_\l(\pi), V_\l(\Pi)$ depend on $\l$, and the best equilibrium for $V_\l$ or $V_\l(\pi)$ may also depend on $\l$. 

(ii) The instability considered in the paper is due to the equilibria. Since $\l$ does not affect the equilibria, so the aggregated optimal game values $V_\l, V_\l(\pi), V_\l(\Pi)$ are actually stable in terms of $\l$. In fact, if we assume $\cR$ is bounded, then obviously $V_\l, V_\l(\pi), V_\l(\Pi)$ are uniformly Lipschitz continuous in $\l$.
\end{rem}

\begin{eg}
\label{eg-lamda1}
For the example in Table \ref{tab:illustrative}, noting that $\l_2=1-\l_1$, we have
\beaa
&\dis \wh V_\l = \max\Big(100, ~102 \l_1, ~102 \l_2\Big) = \max\Big(100, ~102 \l_1, ~102 (1-\l_1)\Big),\\
&\dis V_\l = \max\Big(\l_1 + 2\l_2,~ 3 \l_1 + \l_2\Big) = \max\Big(2-\l_1,~ 1+2\l_1\Big),
\eeaa
and thus
\beaa
E_\l := {V_\l\over \wh V_\l} = \left\{\ba{lll} {2-\l_1\over 102(1-\l_1)}, ~ 0<\l_1 < {1\over 51};\ms\\ {2-\l_1\over 100} , ~  {1\over 51}\le \l_1 <{1\over 3};\ms\\  {1+2\l_1\over 100} , ~  {1\over 3}\le \l_1 <{50\over 51};\ms\\ {1+2\l_1\over 102\l_1} , ~  {50\over 51}\le \l_1 <1. \ea\right.
\eeaa
We note that, the optimal control for $\wh V_\l$ and the best equilibrium for $V_\l$ may depend on $\l$ and have jumps at $\l_1 = {1\over 51}, {1\over 3}, {50\over 51}$, but the values $\wh V_\l, V_\l, E_\l$ are (Lipschitz) continuous in $\l$.
\end{eg}

\section{A dynamic game}
\label{sect-dynamic}
\setcounter{equation}{0}
In this section we consider a continuous time model. Let $(\O, \cF, \dbF, \dbP)$ be a filtered probability space, $B$ a $d$-dimensional Brownian motion. We shall consider $N$-player games with drift controls only and thus use weak formulation. Set
\beaa
X\equiv B.
\eeaa
Let $A=A_1\times \cds\times A_N$ be the control set, and $\cA = \cA_1\times \cds\times \cA_N$ be the admissible controls which are $\dbF^X$-progressively measurable and $A$-valued process $\a=(\a^1, \cds, \a^N)$. Define
\bea
\label{cont-J}
\left.\ba{c}
\dis J_i(\a) = \dbE^{\dbP^\a}\Big[g_i(X_T) + \int_0^T f_i(t, X_t, \a_t)dt\Big], \q i=1,\cds, N,\\
\dis \mbox{where}\q
dP^\a = M^\a_T d\dbP,~ M^\a_T = \exp\Big(\int_0^T b(t, X_t, \a_t) \cd dB_t - {1\over 2}\int_0^T |b(t, X_t, \a_t)|^2dt\Big).
\ea\right.
\eea
In this section the following assumption will always be in force, thus $J_i(\a)$ is well defined.

\begin{assum}
\label{assum-dynamic}
 $(b, f): [0, T]\times \dbR^d \times A\to (\dbR^d, \dbR^N)$ is progressively measurable, bounded, and continuous in $a$; and $g: \dbR^d\to \dbR^N$ is measurable and bounded.
\end{assum}
We emphasize that the boundedness of $f$ and $g$ are just for simplicity and can be replaced with appropriate integrability conditions. 

Let $\cE$ denote the set of equilibria $\a^*\in \cA$ in the obvious sense:  
\bea
\label{NE2}
J_i(\a^*) \ge J_i(\a^{*,-i}, \a_i),\q \forall \a_i\in \cA_i, \forall i.
\eea
As in \cite{FRZ}, we define the raw set value $\dbV_0$ of the game as follows:
\bea
\label{dbV0}
\dbV_0 := \big\{J(a^*): a^*\in \cE\big\}\subset \dbR^N.
\eea
As explained in \cite{FRZ}, the main motivation for studying the set value is that it allows one to study the game value dynamically. In the meantime, we have the following obvious equality:
\bea
\label{dbV0duality}
\sup\big\{\ol J(\a^*): \a^*\in \cE\big\} = \sup\big\{{1\over N} \sum_{i=1}^N y_i: y\in \dbV_0\big\}.
\eea
That is, once we understand $\dbV_0$, then the above optimization problem becomes trivial. However, technically it is not convenient to study the raw set value $\dbV_0$. Note that the admissible control set $\cA$ is not compact (not to mention it is finite), then the existence of equilibria may not be trivial, and more seriously it is hard to establish the regularity as in Theorem \ref{thm-mon} (ii). 

On the other hand, for control problems (with $N=1$), the value function is defined as
\beaa
v := \sup_\a J(\a) = \lim_{\e\to 0} J(\a^\e),
\eeaa
where $\a^\e$ is an $\e$-optimal control, rather than $v := J(\a^*)$, where $\a^*$ is an optimal control, which may not exist. 
Motivated by this, we modify $\dbV_0$ and define the set value of the game. For $\e>0$, we say $\a^\e\in \cA$ is an $\e$-equilibrium, denoted as $\a^\e\in \cE^\e$, if 
\bea
\label{eNE}
J_i(\a^\e) \ge J_i(\a^{\e,-i}, \a_i) -\e,\q \forall \a_i\in \cA_i, \forall i.
\eea
As in \cite{FRZ}, we define the set value $\dbV$ of the game as follows:
\bea
\label{dbV}
\dis \dbV := \bigcap_{\e>0} \dbV_\e,\q \dbV_\e := \Big\{y\in \dbR^N: |y-J(\a^\e)|\le \e,~\mbox{for some}~\a^\e\in \cE^\e\Big\}\subset \dbR^N.
\eea

\begin{rem}
\label{rem-setvalue}
(i)  For fixed $\e>0$, the $\e$-equilibrium can be interpreted as bounded rationality, while the true equilibrium corresponds to unbounded rationality. Therefore, the values in $\dbV$ can be viewed as the values with asymptotically unbounded rationality. 

(ii) The set value $\dbV$ also enjoys many other properties, such as regularity/stability, and it is  possible that $\dbV\neq \emptyset = \dbV_0$. In fact, $\dbV\neq \emptyset$ if and only if $\cE^\e\neq\emptyset$ for all $\e>0$. We refer to \cite{FRZ} for more details.

(iii) In general $\dbV \neq \big\{J(\a^*): \a^*\in \cap_{\e>0} \cE^\e\big\}$. In fact, $ \cap_{\e>0} \cE^\e = \cE$ and thus the right side above is exactly $\dbV_0$. Noting again that $\cA$ is not compact, it is possible that $\{\a^\e\}$ do not converge but $\{J(\a^\e)\}$ converge, and the latter limits are the elements of $\dbV$.  
\end{rem}

From now on, we assume $\dbV \neq \emptyset$, which again is a weaker assumption than $\cE\neq\emptyset$. Fix the weights $\l$ as in the previous section. We then define the optimal aggregated game value:
\bea
\label{cont-Vlambda}
 V_\l := \sup_{y\in \dbV} \sum_{i=1}^N\l_i y_i = \lim_{\e\to 0} \sup_{a^\e\in \cE_\e} \sum_{i=1}^N \l_i J_i(a^\e).
\eea

In this setting, a mechanism scheme $\Pi$ consists of $\pi=(\tilde \pi, \check\pi )$, where $\tilde \pi:  C([0, T]; \dbR^d)\times \dbR^N \to \dbR^N$ and $\check\pi : [0, T]\times C([0, T]; \dbR^d)\times \dbR^N \times A\to \dbR^N$ are progressively measurable and $\check\pi $ is adapted. Given $\pi\in \Pi$, we consider the game:
\bea
\label{cont-Jpi}
J^\pi_i(\a) = \dbE^{\dbP^\a}\Big[g_i(X_T) + \tilde \pi_i(X_\cd, g(X_T)) + \int_0^T\!\!\! \big[f_i(t, X_t, \a_t)+ \check\pi _i(t, X_\cd, f(t, X_t, \a_t), \a_t)\big]dt\Big].
\eea
Note that here we allow $\pi$ to depend on the paths of $X$. 
Let $\cE_\pi^\e$ denote the set of $\e$-equilibria of the game with payoff $J^\pi$, we then define
\bea
\label{cont-Vl}
V_\l(\pi) := \lim_{\e\to 0} \sup_{\a^\e\in \cE_\pi^\e} \sum_{i=1}^N \l_i J_i(\a^{\e}),\q V_\l(\Pi) := \sup_{\pi\in \Pi} V_\l(\pi).
\eea
Extend the Hausdorff distance \reff{dPi} in the obvious manner. Theorem \ref{thm-mon}  remains true.
\begin{thm}
\label{thm-moncont}
(i) $V_\l$ is increasing in $\Pi$ in the sense that: 
\beaa
V_\l(\Pi_1) \le V_\l(\Pi_2),\q\mbox{whenever}\q \Pi_1 \subset \Pi_2.
\eeaa

(ii) $V_\l$ is upper semi-continuous on compact $\Pi$  in the sense that:  
\beaa
\limsup_{n\to\infty} V_\l(\Pi_n) \le V_\l(\Pi), \q\mbox{whenever}\q \lim_{n\to\infty} d(\Pi_n, \Pi) =0.
\eeaa
 In particular, if $\Pi$ is compact, $\Pi_n \downarrow \Pi$ and $d(\Pi_n, \Pi) \to 0$, then $V_\l(\Pi_n) \downarrow E_\l(\Pi)$.
\end{thm}
The proof is essentially the same as Theorem \ref{thm-mon}, except that $\cA$ is not finite anymore. We sketch a proof in Appendix. 

\begin{rem}
\label{rem-Picompact}
The assumption that $\Pi$ is also a strong requirement. However, this can be achieved by assuming certain uniform regularity on $\pi$. For example, consider $\Pi = \{\pi = (\tilde \pi, \check\pi )\}$ such that $\check\pi  \equiv 0$, $\tilde \pi_i= \tilde \pi_i(g(X_T))$, and $\tilde\pi_i$ is uniformly Lipschitz continuous with a fixed modulus of continuity function $\rho$ and $|\tilde\pi(0)|\le C_0$. Then $\Pi$ is compact. 
\end{rem}

We now introduce the $\k$-implementation and the taxation mechanism in this setting. We consider only the full information case, and one can easily  restrict it to the partial information case as well. 
\begin{eg}
\label{eg-cont1}
(i) Given $\k = (\tilde \k, \check\k ) \in [0, \infty)^2$,  $\Pi_\k$ denotes the set of functions $\pi = (\tilde \pi, \check\pi )$:
\beaa
\tilde \pi_i \ge 0,\q\sum_{i=1}^N \tilde \pi_i \le \tilde \k;\qq \check\pi _i \ge 0,\q \sum_{i=1}^N \check\pi _i \le \check\k .
\eeaa

(ii) For $\th = (\tilde \th, \check\th )\in [0, 1]^2$, $\Pi_\th$ denotes the set of functions $\pi = (\tilde \pi, \check\pi )$ such that, there exist functions $\psi = (\tilde \psi, \check\psi )$ satisfying
\beaa
&\dis \tilde \pi(\bx, y) = \tilde \psi(\bx, y) - \tilde \th y,\q \check\pi (t, \bx, y, a) = \check\psi (t, \bx, y, a) - \check\th  y;\\
&\dis  \tilde \psi_i \ge 0,\q\sum_{i=1}^N \tilde \psi_i(\bx, y) \le \tilde \th \sum_{i=1}^N y_i;\q \check\psi _i \ge 0, \q \sum_{i=1}^N \check\psi _i(t, \bx, y, a) \le \check\th  \sum_{i=1}^N y_i.
\eeaa
\end{eg} 

It is interesting and challenging to characterize $V_\l(\Pi_\k)$ and $V_\l(\Pi_\th)$ in this setting. We shall provide a brief discussion in Subsection \ref{sect-PA} below. The following example shows that again $V_\l$ can be discontinuous.
\begin{eg}
\label{eg-cont2}
Let $N=2$, $\l_1 = \l_2 = {1\over 2}$, $A$ as in Section \ref{sect-example}, $g\equiv 0$, $f(t,x,a) = J(a)$ for the $J$ in Table \ref{tab:illustrative}. For $\k = (0, \check\k )$ with $\check\k  \ge 0$,  it is straightforward to show that $E_\l(\Pi_\k) := {V_\l(\Pi_\k)\over \wh V_\l}$ in Example \ref{eg-cont1} (i) is the same as the $E(\check\k )$ in \reff{Ek2}, where $\wh V_\l := \sup_{\a\in \cA} {1\over 2}[J_1(\a) + J_2(\a)]$. Then $E_\l(\Pi_\k)$ and hence $V_\l(\Pi_\k)$ is discontinuous at $\check\k =1$ and $\check\k =4$.
\end{eg} 

\begin{rem}
\label{rem-MFG}
All the results in this section can be extended to mean field games, which consist of infinitely many players. In particular, this is appropriate for social problems where the system is by nature large. However, to avoid the heavy notations, we do not provide details here. We refer to Iseri-Zhang \cite{IZ} for set values of mean field games.
\end{rem}

\section{Further discussions}
\label{sect-further}
\setcounter{equation}{0}
In this section we provide two possible extensions  of our problem.

\subsection{A leader follower problem}
\label{sect-PA}
By nature the problems \reff{Vpi}-\reff{VPi} and \reff{cont-Vl} are leader follower problems with multiple followers, also called Stackelberg games, where the mediator is the leader and the players are  followers. Hence the problem is also intrinsically connected to the principal agent problems with one principal and multiple agents. 

We note that in \reff{cont-Vl} the leader does not have her own interest, she represents the followers' aggregate interests. In particular, as pointed out in Remark \ref{rem-implementation} and Remark \ref{rem-setvalue2}, given the leader's control $\pi\in \Pi$, $V_\l(\pi)$ corresponds to the best equilibrium which is Pareto optimal among the followers' equilibria. It will be very interesting to solve this leader follower problem \reff{cont-Vl}, especially Example \ref{eg-cont1}. However, since $\dbV_\l(\pi)$ has bad stability with respect to $\pi$, in general this is a challenging problem and we leave it for future research.

We may alternatively study the worst equilibrium, corresponding to the price of anarchy:
\bea
\label{Vlworst}
\ul V_\l(\pi) := \lim_{\e\to 0} \inf_{\a^\e\in \cE_\pi^\e} \sum_{i=1}^N \l_i J_i(\a^{\e}),\q \ul V_\l(\Pi) := \sup_{\pi\in \Pi} \ul V^\pi_\l .
\eea
This is the robust approach from the mediator's point of view. The above problem is a max-min problem, however, the min problem is over all (approximate) equilibria, and thus the problem is much more challenging than the standard max-min problems.

Moreover, we can consider more general leader follower problems where the leader has her own interest, or say a utility $J_P(\pi, \a)$. Here for simplicity let's assume the leader's control is still the mechanisms $\pi\in \Pi$. Then correspondingly we can have the following problems corresponding to the worst equilibrium and the best equilibrium, respectively:
\bea
\label{JP}
\left.\ba{c}
\dis \ul V(\pi) := \lim_{\e\to 0} \inf_{\a^\e\in \cE_\pi^\e} J_P(\pi, \a^{\e}),\q \ul V(\Pi) := \sup_{\pi\in \Pi} \ul V(\pi);\\
\dis \ol V(\pi) := \lim_{\e\to 0} \sup_{\a^\e\in \cE_\pi^\e} J_P(\pi, \a^{\e}),\q \ol V(\Pi) := \sup_{\pi\in \Pi} \ol V(\pi).
\ea\right.
\eea
We remark that the (approximate) equilibria are independent of the leader's problem, and thus $\cE_\pi^\e$ is the same as before.
Here $\ul V(\Pi)$ measures the leader's optimal utility in the worst scenario, provided the agents would implement an (approximate) equilibrium. So this is a robust approach for the leader. The problem $\ol V(\Pi)$ corresponding to the best scenario is problematic in practice, however. Unlike in  \reff{cont-Vl}, here the (approximate) optimal equilibrium for the leader, denoted as $\a^*$, may not be Pareto optimal for the followers among all (approximate) equilibria, hence it is hard for the leader to induce the followers to implement $\a^*$.\footnote{We do not have the same concern for $\ul V(\pi)$, because here the leader is just considering the worst scenario, and she has no intention to induce the followers to implement the worst equilibrium.} The problem is closely related to the selection problem, namely given $\pi$, which equilibrium (or even non-equilibrium control) the leader expects the followers will implement. While there are other alternatives, one possibility which sounds reasonable from the practical point of view is to consider the best equilibrium, best for the leader, among all Pareto optimal equilibria, Pareto optimal for the followers. That is, let $\cE^{Pareto}_{\pi,\e}$ denote the set of  $a^\e\in \cE_\pi^\e$ which is $\e$-Pareto optimal in the following sense: there is {\it no} $\tilde a^\e\in \cE_\pi^\e$ such that $J_i^\pi(\tilde \a^\e) \ge  J_i^\pi(\a^\e) + \e$ for all $i$. We then consider
\bea
\label{JPareto}
 V(\pi) := \lim_{\e\to 0} \sup_{\a^\e\in \cE^{Pareto}_{\pi,\e}} J_P(\pi, \a^{\e}),\q V(\Pi) := \sup_{\pi\in \Pi} V^\pi.
\eea
However, it is in general hard to characterize $\cE^{Pareto}_{\pi,\e}$, so \reff{JPareto} could be even more challenging than \reff{JP}  and \reff{cont-Vl}.

\subsection{A central planned economy with large $\th$}
\label{sect-CPE}
The $\th$ in the taxation mechanism can also be interpreted as the portion the mediator, say the government, controls the economy. For example, in a central planned economy, $\th$ is typically large. In particular, when $\th=1$, then in Example \ref{eg-pi} (ii) we have efficiency $E(\th) = 1$, or say the game problem reduces to the control problem. However, for large $\th$, typically the value function $J$ of the base game may depend on $\th$: $J = J(\th, a)$.  Let $\cE^{\th,\e}_\pi$ denote the set of $\e$-equilibria of the mediated game $J^\pi(\th, a)$. Then it will be interesting to consider the following optimization problem, especially when $J$ is decreasing in $\th$:
\bea
\label{CPE}
\sup_{\th\in [0,1]}V_\l(\th) = \sup_{\th\in [0,1]} \sup_{\pi \in \Pi_\th}  V_\l(\pi)  ,\q V_\l(\pi):=    \lim_{\e\to 0} \sup_{\a^\e\in \cE^{\th,\e}_\pi} \sum_{i=1}^N \l_i J_i(\th, \a^{\e}).
\eea
In particular, this could give guideline on how much we would like to have our economy planned or marketized. We remark that,  in this case the value of the optimal control problem also depends on $\th$: 
\bea
\label{CPEhatV}
\wh V_\l(\th) := \sup_{a\in A}\sum_{i=1}^N \l_i J_i(\th, a).
\eea
The the optimization problem \reff{CPE} is not equivalent to maximizing $E_\l(\th) := {V_\l(\th)\over \wh V_\l(\th)}$.

\begin{eg}
\label{eg-CPE}
In the setting of Subsection \ref{sect-theta}, assume $J(\th, a) = (2-\th) J(a)$ is proportional to the $J(a)$ there and $\l_1=\l_2={1\over 2}$. Then one can easily see that $ E(\th)$ is the same as in \reff{Eth1}, and by \reff{optimalV} we have  $\hat V(\th) = 100(2-\th)$. Thus, 
\bea
\label{Eth2}
V(\th)= \left\{\ba{lll} 2(2-\th),\qq 0\le \th < {1\over 103};\\ 51(2-\th),\q~ {1\over 103}\le \th < {1\over 51};\\ 100(2-\th),\q {1\over 51} \le \th \le 1.\ea\right.
\eea
Then, one can easily verify that the optimal $\th$ for \reff{CPE}  is $\th^* = {1\over 51}$.
\end{eg}

The following problem in continuous time model is again very challenging. Let $\th =(\tilde \th, \check\th ) \in [0,1]^2$ and $\a\in \cA$ be as in the setting of Section \ref{sect-dynamic}, define
\bea
\label{cont-Jtha}
\left.\ba{c}
\dis J_i(\th, \a) = \dbE^{\dbP^{\th,\a}}\Big[g_i(\th, X_T) + \int_0^T f_i(t, \th, X_t, \a_t^i)dt\Big], \q i=1,\cds, N,\q\mbox{where}\\
\dis dP^{\th, \a} = M^{\th,\a}_T d\dbP,\q M^{\th, \a}_T = \exp\Big(\int_0^T b(t, \th, X_t, \a_t) dB_t - {1\over 2}\int_0^T |b(t, \th, X_t, \a_t)|^2dt\Big).
\ea\right.
\eea 
Let $\Pi_\th$ be as in Example \ref{eg-cont1} (ii),  for for each $\pi\in \Pi_\th$, define $J^\pi(\th, \a)$ and $\cE^{\th,\e}_\pi$ in the obvious manner. We then have the following optimization problem corresponding to \reff{CPE}:
\bea
\label{CPE2}
\sup_{\th\in [0,1]^2}  \sup_{\pi \in \Pi_\th}  \lim_{\e\to 0} \sup_{\a^\e\in \cE^{\th,\pi}_\e} \sum_{i=1}^N \l_i J_i(\th, \a^{\e}).
\eea
Again we shall leave it for future research.

\section{Conclusion}
\label{sect-conclude}
\setcounter{equation}{0}
It is well known that a non-cooperative game is typically inefficient, in the sense that an equilibrium may have less aggregate payoff than the social optimum. In this paper we study mechanism design by a mediator aiming to improve the efficiency of a game, equivalent to the price of stability concerning the best equilibrium. In particular, we introduce two mechanisms, the $\k$-implementation and the taxation mechanism. The former one contains rewarding only, while the latter one provides both rewarding and punishment.

We focus on the mechanism design with small perturbations of the game. This is possible because the efficiency operator is typically unstable, and thus a small perturbation could improve the efficiency dramatically. The restriction to small perturbation is important in practice, because quite often the mediator has only limited resources (including the right to make certain policy). In particular, when the mediator has several games to mediate but has only limited resources, it could be wise to set a higher priority for her resources  on the game(s) whose efficiency can be improved more dramatically, rather than on the games which are more important for the society as suggested by the conventional wisdom. Moreover, the efficiency operator could be discontinuous at a small parameter, and at those discontinuous points a very small change of the investment can make a big difference on the efficiency. Then it is important to figure out those points, so that the mediator can use her resources in full effect. 

We also investigate the impact of the information on the mediator's ability to improve the efficiency. When the mediator can only provide rewards, we see that the full information and partial information provide the mediator the same power. However, if the mediator is allowed to design punishment mechanisms, she could have larger power to improve the efficiency in the full information case.

However, it is mathematically very challenging to find the optimal mechanism for the mediator, especially in continuous time models. We shall leave them for future research.

\section{Appendix: Proof of Theorem \ref{thm-moncont} (ii)}
\label{sect-appendix}
\setcounter{equation}{0}
We shall only sketch a proof for $\limsup_{n\to\infty} V_\l(\Pi_n) \le V_\l(\Pi)$, where $\Pi_n\to \Pi$ and $\Pi$ is compact. Recall \reff{cont-Vl} and fix an arbitrary $\e>0$. For each $n\ge 1$, there exist $\pi_n =(\tilde \pi_n, \check \pi_n)\in \Pi_n$, $\d_n \le \e$, and $\a^\e_n \in \cE_{\pi_n}^{\d_n}$ such that
\beaa
V_\l(\Pi_n) \le V_\l(\pi_n) +\e   \le \sup_{\a^{\d_n}\in \cE_\pi^{\d_n}} \sum_{i=1}^N \l_i J_i(\a^{\d_n}) + 2\e \le \sum_{i=1}^N \l_i J_i(\a^\e_n) + 3\e. 
\eeaa
By \reff{dPi}, there exists $\pi_n'\in \Pi$ such that $\|\pi_n' - \pi_n\|_\infty \le d(\Pi_n, \Pi) + \e$. Moreover, since $\Pi$ is compact, there exists a subsequence, stilled denoted as $\pi'_n$, such that $\pi'_n \to \pi^*\in \Pi$. Now for $n$ large enough such that $d(\Pi_n, \Pi) \le \e$ and $\|\pi_n'-\pi^*\|_\infty\le \e$, we have 
\beaa
\|\pi_n - \pi^*\|_\infty \le \|\pi_n - \pi_n'\|_\infty + \|\pi_n'-\pi^*\|_\infty \le d(\Pi_n, \Pi) + \e + \|\pi_n'-\pi^*\|_\infty  \le 3\e.
\eeaa 
Recall \reff{eNE}, for any $i$ and $\a_i$ we have
\beaa
J^{\pi^*}_i((\a^\e_n)^{-i}, \a_i) &\le& J^{\pi_n}_i((\a^\e_n)^{-i}, \a_i) + C\|\pi^* - \pi_n\|_\infty \le J^{\pi_n}_i(\a^\e_n) + \d_n + C\|\pi_n' - \pi_n\|_\infty\\
&\le&J^{\pi^*}_i(\a^\e_n)  + C\|\pi_n' - \pi_n\|_\infty +\e \le J^{\pi^*}_i(\a^\e_n)  + C\e,
\eeaa
where $C$ is a generic constant, depending on $T$, which may vary from line to line. 
That is, $\a^\e_n\in \cE_{\pi^*}^{C\e}$. Thus, for $n$ large enough,
\beaa
V_\l(\Pi_n) \le  \sum_{i=1}^N \l_i J_i(\a^\e_n) + 3\e \le \sup_{\a^\e\in \cE_{\pi^*}^{C\e}}  \sum_{i=1}^N \l_i J_i(\a^\e) + 3\e. 
\eeaa
Then, send $n\to\infty$ and $\e\to 0$,
\beaa
\limsup_{n\to\infty} V_\l(\Pi_n) \le \lim_{\e\to 0}  \sup_{\a^\e\in \cE_{\pi^*}^{C\e}}  \sum_{i=1}^N \l_i J_i(\a^\e) = V_\l(\pi^*)\le V_\l(\Pi).
\eeaa

\vspace{-11mm}
\qed

\end{document}